\documentclass[
]{ceurart}

\usepackage{listings}
\begin{document}

\copyrightyear{2025}
\copyrightclause{Copyright for this paper by its authors. Use permitted under Creative Commons License Attribution 4.0 International (CC BY 4.0)}

\conference{SIGIR 2025: Workshop on eCommerce, June 17, 2025, Padua, Italy}

\title{AI Guided Accelerator For Search Experience}

\author[1]{Jayanth Yetukuri}[email=jyetukuri@ebay.com]
\author[1]{Mehran Elyasi}[email=melyasi@ebay.com]
\author[2]{Samarth Agrawal}[email=samagrawal@ebay.com]
\author[1]{Aritra Mandal}[email=arimandal@ebay.com]
\author[1]{Shuang Zhou}[email=shuazhou@ebay.com]
\author[1]{Rui Kong}[email=rukong@ebay.com]

\author[1]{Harish Vempati}[email=havempati@ebay.com]
\author[1]{Ishita Khan}[email=ishikhan@ebay.com]
\address[1]{eBay Inc, San Jose, CA, USA}
\address[2]{eBay Inc, Seattle, WA, USA}

\newcommand{\ebay}{{\emph{eBay}}}
\newcommand{\data}{\mathcal{D}}
\newcommand{\jaccScore}{\emph{Jacc}}
\newcommand{\divFunc}[2]{\mathcal{D}\paranthesis{#1, #2}}

\newcommand{\prob}[1]{\mathbb{P}\left(#1\right)}

\newcommand{\intentDisparity}{intentDisparity}
\newcommand{\trainDiv}{\lambda_{tr}}
\newcommand{\modelDiv}{\lambda_{\theta}}

\newcommand{\classifier}{f}

\newcommand{\model}{{\theta}}
\newcommand{\modelTd}{\theta_{td}}
\newcommand{\modelTdcr}{\theta_{tdcr}}
\newcommand{\modelTdr}{\theta_{tdr}}

\newcommand{\loss}{\mathcal{L}}
\newcommand{\lossRef}{\mathcal{L}_{Ref}}
\newcommand{\lossDiv}{\mathcal{L}_{Div}}

\newcommand{\argmax}[1]{{argmax}_{#1}} 
\newcommand{\argmin}[1]{{argmin}_{#1}} 

\newcommand{\query}{\mathnormal{qi}}
\newcommand{\queryOne}{\mathnormal{q1}}
\newcommand{\queryTwo}{\mathnormal{q2}}
\newcommand{\src}{q}
\newcommand{\tgt}{\mathnormal{t}}
\newcommand{\tgtOne}{\mathnormal{t1}}
\newcommand{\tgtTwo}{\mathnormal{t2}}
\newcommand{\tgtI}{\mathnormal{ti}}
\newcommand{\tgtJ}{\mathnormal{tj}}
\newcommand{\reform}{\mathnormal{r}}
\newcommand{\reformI}{ri}
\newcommand{\reformJ}{\mathnormal{rj}}
\newcommand{\reformOne}{r1}
\newcommand{\reformTwo}{r2}

\newcommand{\paranthesis}[1]{\left(#1\right)}
\newcommand{\squareParanthesis}[1]{\left[#1\right]}

\newcommand{\update}[1]{{#1}}
\newcommand{\remove}[1]{}
\newcommand{\ishita}[1]{\textbf{\color{red}(Ishita: #1)}}
\newcommand{\samarth}[1]{\textbf{\color{blue}(Samarth: #1)}}
\newcommand{\aritra}[1]{\textbf{\color{green}(Aritra: #1)}}
\newcommand{\shuang}[1]{\textbf{\color{cyan}(Shuang: #1)}}
\newcommand{\rui}[1]{\textbf{\color{pink}(Rui: #1)}}
\newcommand{\mehran}[1]{\textbf{\color{orange}(Mehran: #1)}}
\begin{abstract}
Effective query reformulation is pivotal in narrowing the gap between a user’s exploratory search behavior and the identification of relevant products in e-commerce environments. While traditional approaches predominantly model query rewrites as isolated pairs, they often fail to capture the sequential and transitional dynamics inherent in real-world user behavior. In this work, we propose a novel framework that explicitly models transitional queries—intermediate reformulations occurring during the user's journey toward their final purchase intent. By mining structured query trajectories from eBay’s large-scale user interaction logs, we reconstruct query sequences that reflect shifts in intent while preserving semantic coherence. This approach allows us to model a user’s shopping funnel, where mid-journey transitions reflect exploratory behavior and intent refinement.

Furthermore, we incorporate generative Large Language Models (LLMs) to produce semantically diverse and intent-preserving alternative queries, extending beyond what can be derived through collaborative filtering alone. These reformulations can be leveraged to populate Related Searches or to power intent-clustered carousels on the search results page, enhancing both discovery and engagement. Our contributions include (i) the formal identification and modeling of transitional queries, (ii) the introduction of a structured query sequence mining pipeline for intent flow understanding, and (iii) the application of LLMs for scalable, intent-aware query expansion. Empirical evaluation demonstrates measurable gains in conversion and engagement metrics compared to the existing Related Searches module, validating the effectiveness of our approach in real-world e-commerce settings.
\end{abstract}


\maketitle

\section{Introduction}

Search query reformulation is a critical component in enhancing user experience on e-commerce platforms, facilitating effective transitions from broad, exploratory queries to targeted item discovery. While traditional reformulation techniques have largely focused on optimizing search recall or improving immediate query relevance, they often abstract away the multi-stage nature of real-world shopping journeys. On platforms such as eBay, user behavior is frequently manifested as a sequential progression of queries, where buyers iteratively refine or pivot their search intent through a series of exploratory or transitional queries before ultimately converging on a successful outcome.

Despite the progress in query reformulation, prior research has primarily centered on the identification of source-target query pairs mined from historical logs, or on the generation of refined queries for ambiguous or underperforming input. These approaches commonly employ techniques such as representation learning, intent classification, and neural generative models to enhance relevance and diversity in search outcomes. For example, embedding-based models have been shown to capture latent intent signals by encoding user interactions with retrieved results, while sequence generation methods—including neural machine translation (NMT) frameworks~\cite{yetukuri2024}-have demonstrated efficacy in generating semantically coherent and behaviorally grounded reformulations.

However, a significant limitation of existing methods lies in their treatment of query reformulation as a two-point transformation, neglecting the intermediate transitions that characterize users’ ongoing refinement of intent. These transitional states—often observable as mid-journey rewrites—contain valuable signals about user uncertainty, exploration, and behavioral divergence. Accurately modeling these phases is essential for developing next-generation retrieval systems that are not only reactive to user input but also anticipatory of their evolving intent. By incorporating these intermediate transitions into the reformulation pipeline, we aim to enrich the model’s understanding of query context, yielding more adaptive and intent-aware search experiences.

\begin{figure}
\includegraphics[trim={0 4cm 0 3.5cm}, width=\textwidth]{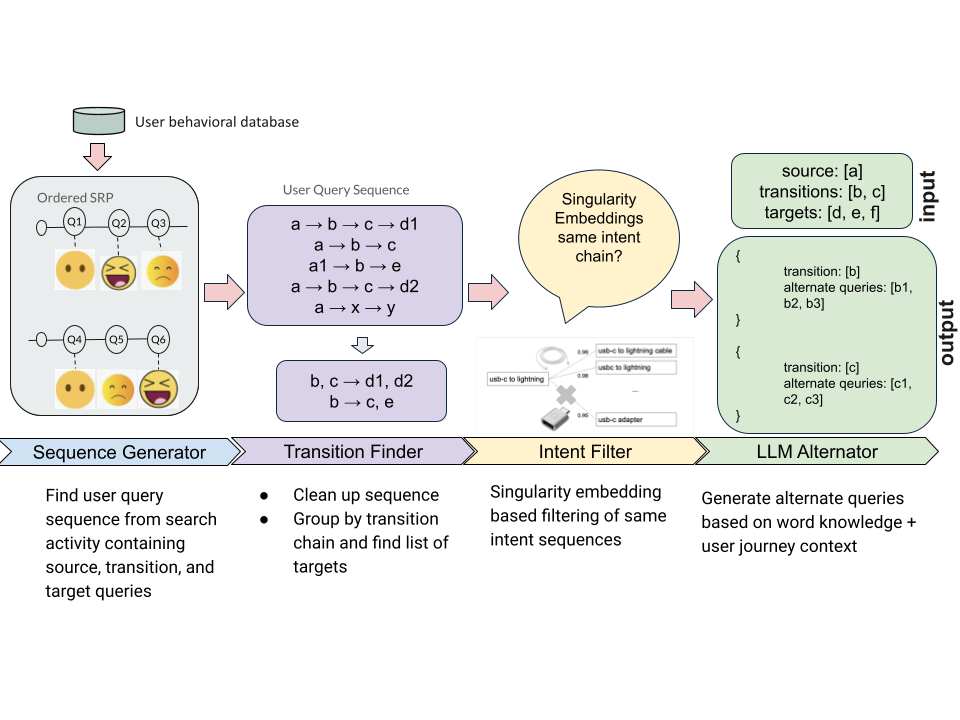}
\caption{High level overview of the AI guided accelerator powering an alternate search grouping}
\label{fig:flow}
\end{figure}

In this work, we propose a novel system that explicitly focuses on identifying and leveraging transitional queries within users’ shopping journeys. By mining structured query sequences from eBay’s behavioral database, our approach captures not just the source and target queries but also the transitional steps users take along the way. Each sequence is organized into three components: a source query, a list of transitional queries, and a list of converging queries. This structured representation allows for a more nuanced understanding of how users refine their searches. To ensure that the mined sequences are effective for training and align with user behavior, we introduce an \textit{intent consistency filtering} mechanism. This step ensures that intent of the source query is preserved throughout the chain by measuring query similarity based on representation learning from item interactions. Sequences failing this test are excluded, improving the quality and reliability of the mined data.

Beyond mining query sequences, our system introduces a novel use of large language models (LLMs) to enhance query reformulation. Specifically, we employ an open-source LLM to generate alternative forms of the converging queries through in-context learning. These alternative queries retain the original intent while introducing diversity, providing a creative yet controlled expansion of the query space. By leveraging LLM-generated alternatives, our approach enriches the user experience and offers greater flexibility in addressing diverse user needs.

The final output of the system is designed to transform the search results page (SRP) by anchoring it around the generated alternative queries. These queries are used to organize the SRP into distinct carousels, each reflecting a different alternative converging query. This structure provides users with an enhanced browsing experience, presenting multiple pathways to explore relevant items while maintaining intent alignment.

In summary, our work advances the state of query reformulation by introducing key innovations: the explicit identification of transitional queries, structured mining of query sequences with intent consistency checks, and the use of LLMs for generating diverse reformulations. These contributions represent a significant departure from traditional methods that primarily focus on source-target pairs. By combining insights from behavioral data with advanced generative techniques, our approach offers a scalable and effective solution for enriching e-commerce search experiences.

\subsection{Related Work}


The optimization of search engines using clickthrough data has been an essential area of research. \cite{joachims2002optimizing} leveraged user clickthrough behavior to enhance query reformulation and improve search relevance. Their approach laid the groundwork for using implicit feedback for query ranking optimization. Furthering this, \cite{boldi2011query} explored query chains, phrase expansions \cite{psimbert}, focusing on the patterns and models of query reformulation derived from user behavior. These studies provide foundational insights into the mechanisms of query transition and user intent modeling.

Representation learning plays a pivotal role in modern query understanding and similarity measurement. \cite{mikolov2013distributed} introduced distributed word representations, a widely adopted foundational method in search and information retrieval tasks. \cite{saadany-etal-2024-centrality}\cite{10.1145/3627673.3679080} further explore training joint representations using centrality scores for product retrieval. Extending these ideas, Ai et al. (2016) proposed a deep listwise context model that incorporates user behavioral signals, demonstrating its effectiveness in ranking refinement tasks. These advances highlight the evolution of representation learning from static embeddings to dynamic, context-aware models.

The emergence of large language models (LLMs) has revolutionized query generation and understanding. \cite{brown2020language} introduced GPT-3, showcasing the potential of in-context learning for few-shot tasks. This capability has become central to the application of LLMs in generating alternative query forms. Additionally, Scao et al. (2022) introduced BLOOM, an open-access multilingual LLM, emphasizing the growing accessibility of these powerful models for research and practical applications in query reformulation.

E-commerce search presents unique challenges for query understanding. \cite{dai2024enhancing} proposed query rewriting techniques tailored for e-commerce platforms, utilizing reinforcement learning to enhance search relevance. Complementing this, Li et al. (2017) explored session-based intent modeling using multi-task learning, demonstrating how user sessions can reveal nuanced intents. These works underscore the importance of domain-specific adaptations for effective query understanding.

Diversity in search results is a critical aspect of user satisfaction. \cite{carbonell1998use} introduced the Maximum Marginal Relevance (MMR) framework, a method to ensure diversity-based reranking of search results. This approach is particularly relevant for structuring search results pages (SRPs) in carousels. Additionally, \cite{liu2010personalized} explored the organization of personalized news recommendations based on click behavior, offering insights into user-centric result anchoring methods. These studies contribute to optimizing search results for both relevance and diversity.

\section{Architecture}

\begin{figure*}
\includegraphics[trim={0 1cm 0 0}, width=\textwidth]{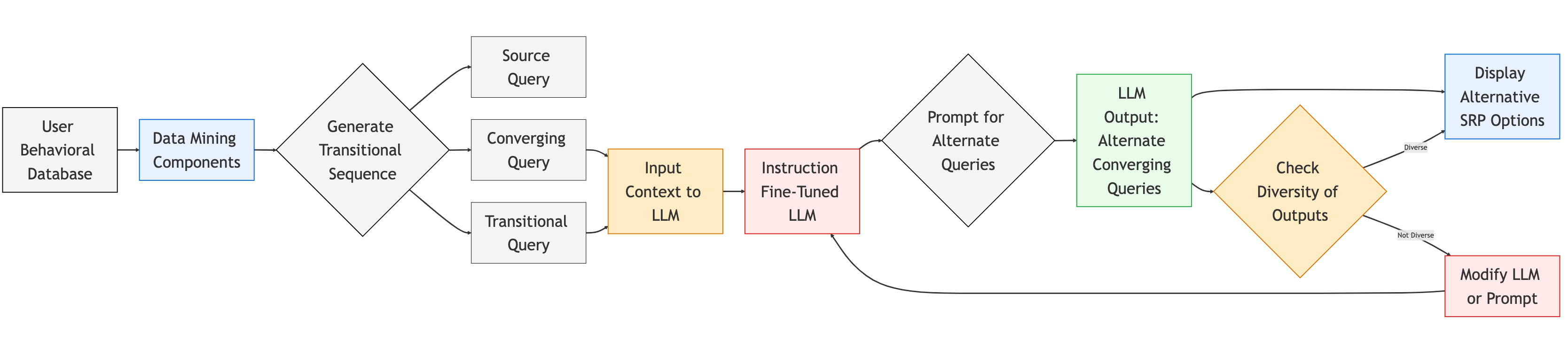}
\caption{High level overview of the AI guided accelerator powering an alternate search grouping}
\label{fig:llmflow}
\end{figure*}

The high-level architecture of the proposed system is illustrated in Figure~\ref{fig:flow}. Our pipeline begins by mining structured query sequences from eBay's large-scale behavioral logs. Each query sequence is segmented into three core components: a source query, a list of transitional queries, and a set of converging (target) queries that reflect successful product discovery.

We introduce a dedicated Transition Finder module that aggregates and refines these raw sequences, applying a series of normalization and de-duplication heuristics to produce a cleaned dataset suitable for downstream processing. Following this, an Intent Filtering step ensures semantic coherence across the query chain. Specifically, we filter out sequences in which the core user intent—anchored in the source query—is not preserved throughout the transitions. This step leverages a learned similarity metric between query representations, where queries are embedded based on the set of items retrieved for each, following techniques outlined in~\cite{Mandal2019QueryRU}.

The filtered, intent-consistent query sequences are then passed through an LLM Alternator module. This component uses in-context learning with a high-capacity open-source LLM~\cite{kim2024solar107bscalinglarge} to generate semantically aligned but lexically diverse alternative converging queries. These alternatives are intended to augment the original converging queries and introduce novel yet relevant reformulations.

The final output of this pipeline powers an enhanced search experience by surfacing these LLM-generated alternate queries as anchor points in carousels on the Search Results Page (SRP). This not only diversifies user exploration pathways but also aligns with latent user intent, offering a more engaging and efficient discovery process.

\section{Methodology}
\subsection{Sequence Generator and Transition Finder}

This study focuses on analyzing user sessions to identify key actions within e-commerce platforms, specifically targeting events categorized as ``bbowac'' (buy, bid, offer, watch, ask, cart click). The data collection process involves analyzing user search activities, which includes the search query, results they interact with, and any subsequent actions they take. This stage aims to identify the longest chains of transitional queries within user sessions. Each session must contain at least one ``bbowac'' event and two queries. Consider a session sequence such as \( a \rightarrow b \rightarrow c \rightarrow d \, (\text{bbowac}) \rightarrow e \rightarrow f \, (\text{bbowac}) \). In this sequence, the event at \( d \) is a bbowac event, prompting the sequence to be split into \( a \rightarrow b \rightarrow c \rightarrow d \) and \( e \rightarrow f \). This results in the generation of two separate sequences: \( a \rightarrow b \rightarrow c \rightarrow d \) and \( e \rightarrow f \). This approach ensures that the data captures meaningful transitional query chains.

\subsection{Intent Filter}
The intent filter process aims to establish an intent boundary within a sequence of queries during a single search session. This is achieved by employing an independent query similarity model \cite{mandal2023semantic} to evaluate the similarity between successive queries. The process involves traversing the query sequence in reverse, beginning from a converting query, and continuing until a query is encountered whose similarity to the preceding query falls below a predetermined threshold. This threshold serves as a hyperparameter within the application. Consider the following example of a sequence of queries with corresponding similarity scores:
\begin{itemize}
    \item macbook [0.6] -> iphone 12 [0.9] -> iphone 12 red [0.95] -> and iphone 12 128gb.
\end{itemize}

Based on these similarity scores, the updated sequence would be:
\begin{itemize}
     \item iphone 12 → iphone 12 red → iphone 12 128gb.
\end{itemize}

The query `macbook' is excluded as it lies outside the intent boundary, decided by domain experts.

\subsection{LLM Alternator}
The purpose of the LLM component in the overall flow is to find query alternatives, i.e., potential converging queries for an input transitional query from the perspective of world knowledge, while using the user transitional data mined from the previous components as an input context fed to the LLM. We performed instruction fine tuning on an open source LLM from hugging face (Solar-10B-Instruct) ~\cite{kim2024solar107bscalinglarge}, where the full user journey containing a source query, a list of transitional queries, and a list of converging queries, mined from the user behavioral database at ebay is provided as input to the LLM. The instruction prompt to the LLM is written to guide the LLM to generate a list of alternate converging queries that are different form the input list of converging queries, and can be provided to the e-commerce user whenever they land on a Search Results Page (SRP) with any of the input transitional queries. These alternate queries should be relevant and have a high likelihood of leading to a transaction. The instruction fine tuning was done based on few shot examples on queries where we gathered a few user journey based query transition chain examples. To illustrate the functionality of our system, we present an example mined from user behavioral logs. The extracted query sequence is structured as follows:

\begin{itemize}
    \item \textbf{Source Queries:} [``18k gold dacid yarmen'', ``david yurman chain gold 18k'']
    \item \textbf{Transitional Query:} [``18k gold diamonds necklace'']
    \item \textbf{Converging Queries:} [``david yurman chain gold 18k'', ``18k gold david yurman'']
\end{itemize}

We use this transitional query as input to our LLM-based alternator module, guided by a carefully constructed in-context learning prompt. The LLM is tasked with generating a set of semantically relevant, yet non-redundant, alternate converging queries that are aligned with the original user intent but exclude any of the mined converging queries. The resulting LLM output is as follows:

\begin{quote}
\texttt{%
\hspace*{1em}``transitional query'': ``18k gold diamonds necklace'',\\
\hspace*{1em}``alternate queries'': [``18k white gold diamond necklace'', ``18k yellow gold diamond necklace'', ``18k gold diamond pendant necklace'', ``18k gold diamond necklace david yurman'', ``18k gold diamond necklace tiffany \& co'', ``18k gold diamond necklace cartier'', ``18k gold diamond necklace van cleef \& arpels'']%
}
\end{quote}

This output demonstrates the model’s ability to preserve the semantic intent of the transitional query while introducing variation in product attributes (e.g., gold type, pendant structure, brand specificity). Such high-quality, brand- and category-aware alternatives are well-suited for downstream use in search experience modules such as dynamic SRP carousels or suggestion panels. Our observations from the LLM-generated outputs reveal several encouraging trends:

\begin{enumerate} 
    \item The LLM adheres closely to the intended problem formulation, producing a set of alternative queries corresponding to each transitional query input. In the illustrative example considered, the input sequence contains a single transitional query, and the model appropriately returns multiple semantically aligned alternatives.
    \item The model successfully respects the constraint of avoiding verbatim repetition of user-generated converging queries. This promotes diversity in the candidate pool while maintaining alignment with the original intent trajectory.
    \item Notably, the LLM demonstrates strong capability in extracting salient query aspects, such as brand, from the input context and effectively propagating them through the generated alternatives. This reflects a nuanced understanding of e-commerce-specific attributes and contextual anchoring.
    \item The resulting alternative queries are not only coherent but also contextually appropriate for deployment as anchors in a modified Search Results Page (SRP). They serve as viable suggestions for downstream recommendation modules or carousel-based experiences, enriching the user journey with meaningful query reformulations.
\end{enumerate}

The operational flow of this generative process is depicted in Figure~\ref{fig:llmflow}, illustrating the LLM’s role in enhancing transitional state modeling within our query rewriting pipeline. One of the challenges we addressed at this step is inclusion of diversity among the converging queries. Indeed, the search experience becomes redundant and unattractive when the resulting alternative query groups are not diverse among themselves in context to the original source query. One such example output that lacks diversity from LLM alterations is the following: \textit{output}: [{``transitional query'': ``air jordan 3'', ``alternate queries'': [``air jordan 3 retro'', ``air jordan 3 black cement'', ``air jordan 3 true blue'', ``air jordan 3 history of flight'', ``air jordan 3 fire red'', ``air jordan 3 white cement'', ``air jordan 3 university blue'']} {``transitional query'': ``airpods pro case'', ``alternate queries'': [``airpods pro leather case'', ``airpods pro silicone case'', ``airpods pro hard case'', ``airpods pro wallet case'', ``airpods pro transparent case'']}].


\begin{figure*}
\includegraphics[trim={5cm 2cm 0 1cm}, width=0.9\textwidth]{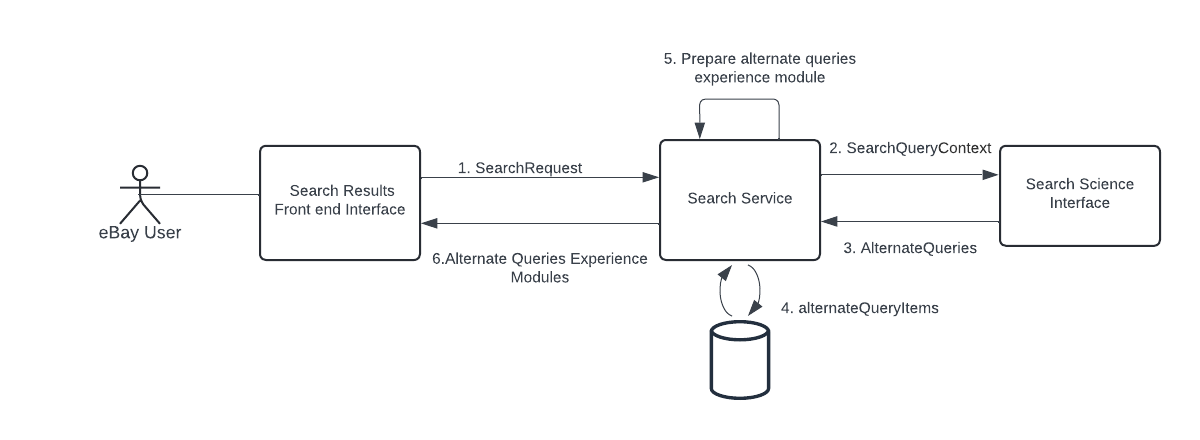}
\caption{Schematic diagram showing using search services with alternate query experience}
\label{fig:application1}
\end{figure*}

\subsection{Applications}
\begin{figure*}
\includegraphics[trim={0cm 0cm 0 0cm}, width=\textwidth]{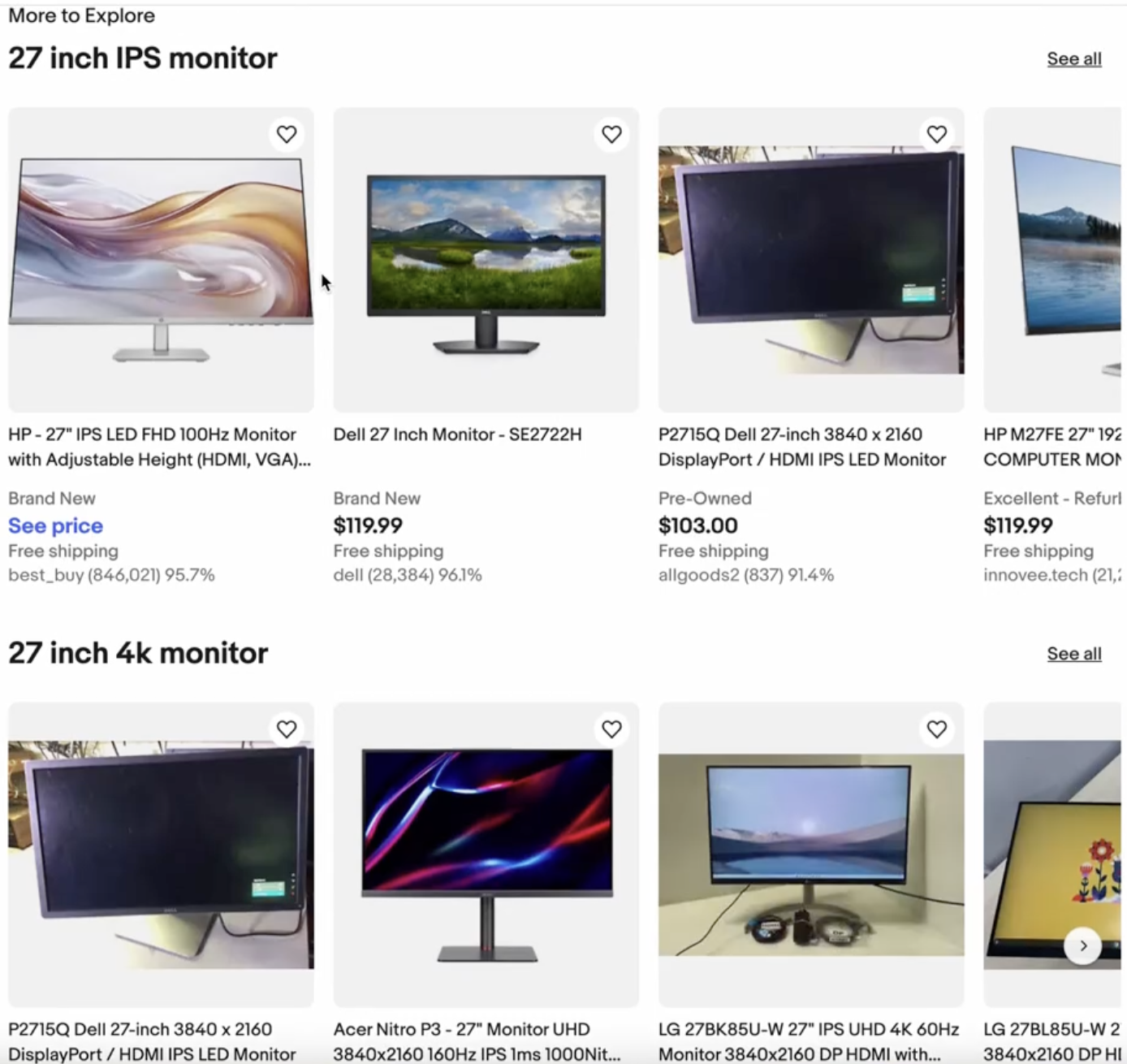}
\caption{Alternate queries for 27 inch monitor input query}
\label{FIX:MODULE}
\end{figure*}

\paragraph{Alternate Search Experience Module:}
The is a high level interaction model from Search Services for alternate queries experience is shown in Figure ~\ref{fig:application1}. The search request is sent from search front end services. Search services initiate the request to search sciences passing along the request and additional related context. Search sciences responds back with suggested alternatives for the queries based on the science models shared above. The items related with alternate queries are further organized and enriched and an experience module is created. The alternate queries experience module as shown in Fig~\ref{FIX:MODULE} is delivered to the front end services.

\paragraph{Related Searches Module:}
The ``Related Searches'' feature on eBay is designed to enhance the user experience by providing a list of search terms or products related to a user's original query \cite{10.1145/1935826.1935927}. This feature helps users refine their searches, explore related topics, and discover new items of interest. Related searches are typically displayed as clickable links or suggested search terms on a search results page, which can improve user experience and drive conversions. For example, if a user searches for ``iphone case'', the Related Searches module might suggest terms like ``iphone magsafe case'' or ``iphone 11 case''. Similarly, a search for smart watch" might yield suggestions such as "smart watch samsung" or "apple watch."

\section{Evaluation}

\begin{table}
  \caption{Relative performance improvements}
  \label{tab:freq}
  \begin{tabular}{lrr}
    \toprule
     & Click-through rate  & Conversions \\
    \midrule
    Intent filtered & -33.6\% & -40.3\% \\
    LLM Alternator  & +32.2\% & +38.3\% \\
  \bottomrule
\end{tabular}
\end{table}

We evaluate the effectiveness of our proposed approach in the context of the Related Search (RS) task, using click-through rate (CTR) and conversion rate as key proxies for assessing recommendation quality. Our evaluation is structured around two stages in the pipeline: (i) output from the system up to the Intent Filtering stage (as depicted in Figure~\ref{fig:flow}), and (ii) output from the LLM Alternator, which generates additional converging queries through in-context learning. To benchmark performance, we conduct a comparative analysis against the production RS system deployed at eBay, using a two-week lookback window of anonymized user interaction data. To maintain user privacy and data compliance, we report only the relative change in conversion rates with respect to the baseline.

As shown in Table~\ref{tab:freq}, suggestions based solely on mined sequences up to the intent filtering step tend to underperform. This is largely attributed to a lack of lexical and semantic diversity, resulting in limited novelty in user-facing recommendations. In contrast, augmenting the RS candidates with LLM-generated alternatives significantly improves both CTR and conversion rates, outperforming the production system. This improvement highlights the utility of LLMs in enhancing the expressiveness and intent alignment of query suggestions while maintaining relevance.

These results demonstrate the importance of combining structured behavioral mining with generative query expansion to balance precision with diversity in large-scale search recommendation systems.

\section{Conclusion and Future Work}

This study presents a novel approach to enhancing e-commerce item retrieval by leveraging transitional buyer queries, capturing the intermediate stages of a shopper's journey. By mining sequential query patterns from user behavior logs and applying large language models (LLMs) to generate diverse, semantically aligned reformulations, we address a key challenge in query understanding—bridging the gap between initial exploratory searches and purchase-driven intent.

Our framework demonstrates that integrating LLMs into the query reformulation pipeline enables scalable, intent-aware retrieval mechanisms that adapt to evolving user goals. The LLM-generated alternate queries serve not only to expand the query space but also to personalize and deepen search experiences by surfacing more contextually relevant results.

This work represents a foundational step toward operationalizing LLMs within real-world e-commerce search systems, highlighting their potential in query expansion, intent preservation, and search journey optimization. Future work will focus on improving alignment via prompt tuning, real-time inference optimization, and hybrid approaches that combine structured behavioral mining with generative models for even more nuanced and actionable search personalization. 

\section*{Declaration on Generative AI}

During the preparation of this work, the author(s) used ChatGPT (GPT-4) and Grammarly in order to: Grammar and spelling check. After using these tool(s)/service(s), the author(s) reviewed and edited the content as needed and take(s) full responsibility for the publication's content.

\bibliography{references}

\end{document}